\documentclass[showpacs,10pt,twocolumn,prb]{revtex4-1}

\usepackage{amsmath}
\usepackage{amssymb}
\usepackage{graphicx}
\usepackage{amssymb}
\usepackage{graphics}
\usepackage{epsfig}
\usepackage{CJK}
\usepackage{color}

\begin{document}

\begin{CJK*}{GBK}{Song}
\title{Anisotropic magnetic entropy change in Cr$_2$X$_2$Te$_6$ (X = Si and Ge)}
\author{Yu Liu and C. Petrovic}
\affiliation{Condensed Matter Physics and Materials Science Department, Brookhaven National Laboratory, Upton, New York 11973, USA}
\date{\today}

\begin{abstract}
Intrinsic, two-dimensional (2D) ferromagnetic semiconductors are an important class of materials for spintronics applications. Cr$_2$X$_2$Te$_6$ (X = Si and Ge) semiconductors show 2D Ising-like ferromagnetism, which is preserved in few-layer devices. The maximum magnetic entropy change associated with the critical properties around the ferromagnetic transition for Cr$_2$Si$_2$Te$_6$ $-\Delta S_M^{max} \sim$ 5.05 J kg$^{-1}$ K$^{-1}$ is much larger than $-\Delta S_M^{max} \sim$ 2.64 J kg$^{-1}$ K$^{-1}$ for Cr$_2$Ge$_2$Te$_6$ with an out-of-plane field change of 5 T. The rescaled $-\Delta S_M(T,H)$ curves collapse onto a universal curve independent of temperature and field for both materials. This indicates similar critical behavior and 2D Ising magnetism, confirming the magnetocrystalline anisotropy that could preserve the long-range ferromagnetism in few-layers of Cr$_2$X$_2$Te$_6$.
\end{abstract}
\maketitle
\end{CJK*}

\section{INTRODUCTION}

Layered ferromagnets such as Cr$_2$Ge$_2$Te$_6$, CrI$_3$, and Fe$_3$GeTe$_2$ have recently attracted considerable attention since long-range ferromagnetism (FM) persists in atomically thin devices.\cite{McGuire0, McGuire, Huang, Gong, Seyler, Huang1} Intrinsic magnetic order is not allowed at finite temperature in the two-dimensional (2D) isotropic Heisenberg model by the Mermin-Wagner theorem \cite{Mermin}, however large magnetocrystalline anisotropy in van der Waals (vdW) magnets would lift this restriction.

Bulk Cr$_2$X$_2$Te$_6$ (X = Si and Ge) exhibit FM below the Curie temperature ($T_c$) of 32 K for Cr$_2$Si$_2$Te$_6$ and 61 K for Cr$_2$Ge$_2$Te$_6$, respectively, and show large magnetocrystalline anisotropy as a result of strong spin-orbit coupling (SOC).\cite{Ouvrard, Carteaux1, Carteaux2, Casto, Zhang} Neutron scattering measurements showed that bulk Cr$_2$Si$_2$Te$_6$ is a strongly anisotropic 2D Ising-like ferromagnet with a critical exponent $\beta = 0.17$ and a spin gap of $\sim$ 6 meV.\cite{Carteaux3} On the other hand, recently observed $\beta = 0.151$ and a much smaller spin gap of $\sim$ 0.075 meV argue that the spins in Cr$_2$Si$_2$Te$_6$ are Heisenberg-like.\cite{Williams} Cr$_2$Ge$_2$Te$_6$ is proposed to be a 2D Heisenberg ferromagnet based on spin wave theory,\cite{Gong} but was also found to follow the tricritical mean-field model,\cite{LinGT} calling for further studies. The magnetocaloric effect (MCE) in the FM vdW materials is also of interest since it can give insight into the magnetic properties. Fe$_{3-x}$GeTe$_2$ with $T_c$ = 225 K shows the maximum value of magnetic entropy change $-\Delta S_M^{max}$ about 1.1 J kg$^{-1}$ K$^{-1}$ at 5 T.\cite{Verchenko} CrI$_3$ exhibits anisotropic $-\Delta S_M^{max}$ with values of 4.24 and 2.68 J kg$^{-1}$ K$^{-1}$ at 5 T for $\mathbf{H//c}$ and $\mathbf{H//ab}$, respectively.\cite{YuLIU}

In this work we studied the anisotropic magnetocaloric effect associated with the critical behavior of Cr$_2$X$_2$Te$_6$ (X = Si and Ge) single crystals. The magnetocrystalline anisotropy constant $K_u$ is temperature-dependent, and is evidently larger for Cr$_2$Si$_2$Te$_6$ when compared to Cr$_2$Ge$_2$Te$_6$. The maximum magnetic entropy change in out-of-plane field up to 5 T $-\Delta S_M^{max} \sim$ 5.05 J kg$^{-1}$ K$^{-1}$ for Cr$_2$Si$_2$Te$_6$ is nearly double of $-\Delta S_M^{max} \sim$ 2.64 J kg$^{-1}$ K$^{-1}$ for Cr$_2$Ge$_2$Te$_6$. Critical exponents $\beta$, $\gamma$, and $\delta$ and critical isotherm analysis suggest 2D Ising-like spins. This is further confirmed by the scaling analysis of magnetic entropy change $-\Delta S_M(T,H)$, in which the rescaled $-\Delta S_M(T,H)$ collapse on a universal curve. Our work provides evidence for magnetocrystalline anisotropy that drives the 2D Ising ferromagnetic state in few layers of Cr$_{2}$X$_{2}$Te$_{6}$ (X = Si and Ge).

\section{EXPERIMENTAL DETAILS}

Single crystals of Cr$_2$X$_2$Te$_6$ (X = Si and Ge) were fabricated by the self-flux technique starting from an intimate mixture of pure elements Cr (3N, Alfa Aesar) powder, Si or Ge (5N, Alfa Aesar) pieces and Te (5N, Alfa Aesar) pieces with a molar ratio of 1 : 2 : 6. The starting materials were vacuum-sealed in a quartz tube, heated to 1100 $^\circ$C over 20 h, held at 1100 $^\circ$C for 3 h, and then cooled to 680 $^\circ$C at a rate of 1 $^\circ$C/h. The x-ray diffraction (XRD) data were taken with Cu $K_{\alpha}$ ($\lambda=0.15418$ nm) radiation of a Rigaku Miniflex powder diffractometer. The dc magnetization was collected in Quantum Design MPMS-XL5 system. The magnetic entropy change $-\Delta S_M(T,H)$ from the dc magnetization data was estimated using the Maxwell relation.

\section{RESULTS AND DISCUSSION}

\subsection{Structural and basic magnetization data}

\begin{figure}
\centerline{\includegraphics[scale=1]{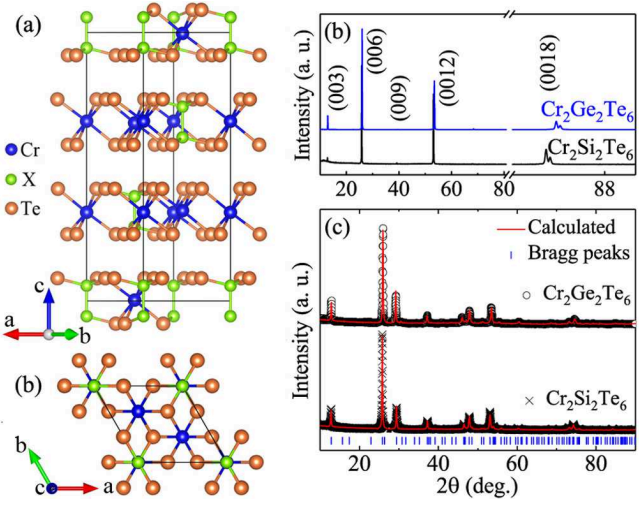}}
\caption{(Color online) Crystal structure of Cr$_2$X$_2$Te$_6$ (X = Si and Ge) from (a) side and (b) top views. (c) Single crystal x-ray diffraction (XRD) and (d) powder XRD patterns of Cr$_2$X$_2$Te$_6$ (X = Si and Ge). The vertical tick marks represent Bragg reflections of the $R\bar{3}h$ space group.}
\label{XRD}
\end{figure}

\begin{figure}
\centerline{\includegraphics[scale=1]{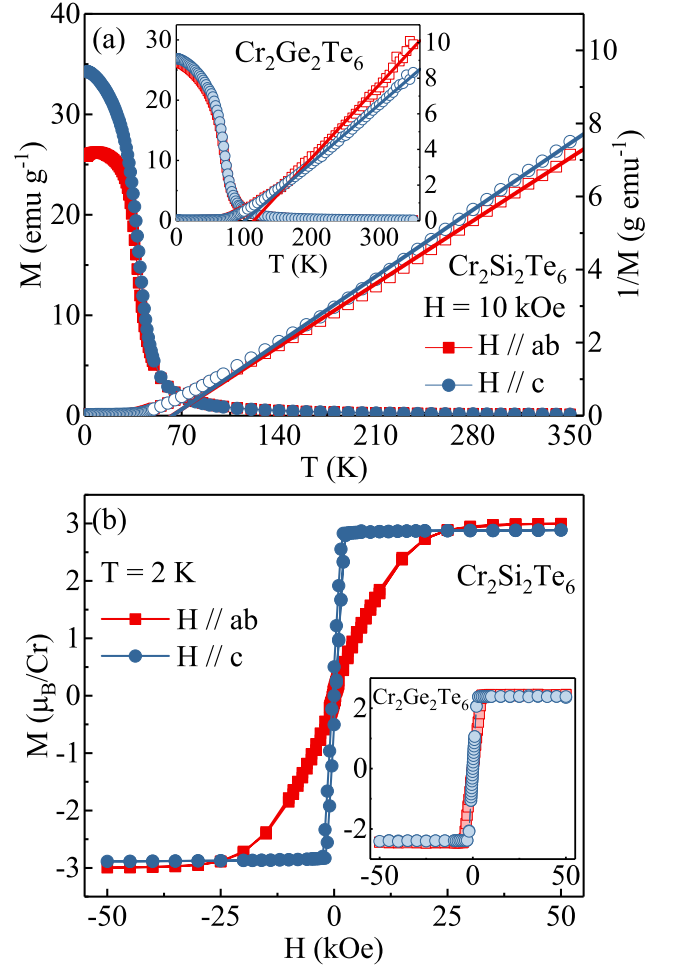}}
\caption{(Color online) (a) Temperature dependence of zero field cooling (ZFC) magnetic susceptibility $\chi$ (left axis) and corresponding $1/\chi$ (right axis) for Cr$_2$X$_2$Te$_6$ (X = Si and Ge) measured in in-plane and out-of-plane field of $H$ = 10 kOe. (b) Field dependence of magnetization measured at $T$ = 2 K.}
\label{MTH}
\end{figure}

\begin{table}
\caption{\label{tab1}The parameters obtained from fits of the $1/M$ vs $T$ data for Cr$_2$X$_2$Te$_6$ (X = Si and Ge) single crystals.}
\begin{ruledtabular}
\begin{tabular}{lllll}
   & Fit $T$ range & $C$ & $\theta_p$ & $\mu_{eff}$\\
   & (K) & (emu K/mol) & (K) & ($\mu_B$/Cr)\\
   \hline
   X = Si & & & &\\
   $\mathbf{H//c}$ & 130 $\leq T \leq$ 300 & 1.70(1) & 63(1) & 3.68(2) \\
   $\mathbf{H//ab}$ & 130 $\leq T \leq$ 300 & 1.60(1) & 63(1) & 3.57(1) \\
   \hline
   X = Ge & & & & \\
   $\mathbf{H//c}$ & 150 $\leq T \leq$ 300 & 1.22(1) & 114(2) & 3.12(1) \\
   $\mathbf{H//ab}$ & 150 $\leq T \leq$ 300 & 1.54(2) & 101(1) & 3.51(2) \\
\end{tabular}
\end{ruledtabular}
\end{table}

Bulk Cr$_2$X$_2$Te$_6$ (X = Si and Ge) were first synthesized by Carteaux \emph{et al.}.\cite{Carteaux1,Carteaux2} They crystalize in a layered structure [Fig. 1(a)]. The Cr ions are located at the centers of slightly distorted octahedra of Te atoms. The short X-X bonds result in X-X dimers forming an ethane-like X$_2$Te$_6$ groups, similar to P-P dimers in CdPS$_3$.\cite{Zhukov} Figure 1(c) presents the single crystal x-ray diffraction (XRD) data. The observed (00l) peaks distinctly shift to higher angles in Cr$_2$Ge$_2$Te$_6$ when compared to Cr$_2$Si$_2$Te$_6$ indicating a smaller vdW gap in Cr$_2$Ge$_2$Te$_6$. The powder XRD data can be indexed in the $R\bar{3}h$ space group [Fig. 1(d)]. The determined lattice parameters are $a = 6.772(2)$ {\AA} and $c = 20.671(2)$ {\AA} for Cr$_2$Si$_2$Te$_6$ [$a = 6.826(2)$ {\AA} and $c = 20.531(2)$ {\AA} for Cr$_2$Ge$_2$Te$_6$], in agreement with the reported values.\cite{Carteaux1,Carteaux2}

Figure 2(a) presents the temperature dependence of the zero field cooling (ZFC) magnetization $M(T)$ measured in $H$ = 10 kOe applied in the $ab$ plane and parallel to the $c$ axis, respectively. The FM transition stems from the near-90$^\circ$ Cr-Te-Cr superexchange interaction and is observed in both materials. An apparent bifurcation at low temperature is observed in Cr$_2$Si$_2$Te$_6$. The absence of bifurction in Cr$_2$Ge$_2$Te$_6$ indicates smaller magnetic anisotropy. The smaller vdW gap and larger in-plane Cr-Cr distance in Cr$_2$Ge$_2$Te$_6$ contribute to the enhancement of the $T_c$ from 32 K for Cr$_2$Si$_2$Te$_6$ to 63 K for Cr$_2$Si$_2$Te$_6$. The $1/M$ vs $T$ curves at high temperature follow the Curie-Weiss law, $\chi(T) = M/H = C/(T-\theta_p)$, where $\chi$ is magnetic susceptibility, $M$ is magnetization, $C$ is the Curie constant, and $\theta_p$ is the Weiss temperature. The obtained parameters $C$ and $\theta_p$ are listed in Table I. The positive Weiss temperatures $\theta_p$, nearly twice the values of $T_c$, for both directions suggest strong short-range FM correlation in Cr$_2$X$_2$Te$_6$ (X = Si and Ge) above $T_c$. The effective magnetic moment $\mu_{eff} \approx \sqrt{8C}$ is also listed in Table I. The values are close to the theoretical value expected for Cr$^{3+}$ of 3.87$\mu_B$. The isothermal magnetization at $T$ = 2 K  is shown in Fig. 2(b). We estimate the saturation magnetization $M_s$ from the intercept of a linear fit of $M(H)$ at high field and the saturation field $H_s$ as the point of deviation from the linear behavior. The derived $M_s \approx$ 2.86(1) $\mu_B$/Cr with out-of-plane field for Cr$_2$Si$_2$Te$_6$ is larger than that of 2.40(2) $\mu_B$/Cr for Cr$_2$Ge$_2$Te$_6$. The saturation field $H_s \approx 3$ kOe with out-of-plane field is smaller than $H_s \approx 5$ kOe with in-plane field and is much smaller than that of 25 kOe for Cr$_2$Si$_2$Te$_6$. These results confirm the easy $c$-axis and smaller magnetic anisotropy in Cr$_2$Ge$_2$Te$_6$, in agreement with previous reports.\cite{Carteaux1,Carteaux2,Casto,Zhang}

\begin{figure}
\centerline{\includegraphics[scale=1]{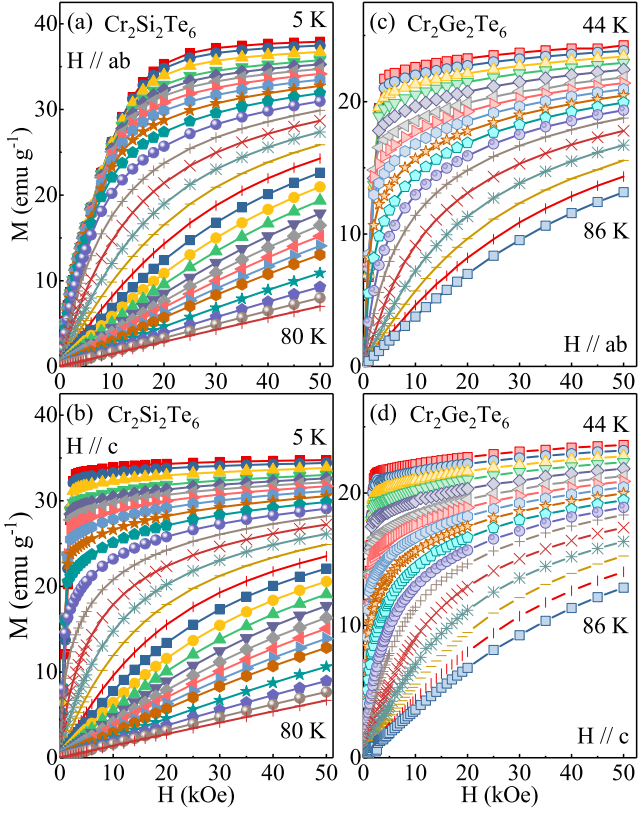}}
\caption{(Color online) Typical initial isothermal magnetization curves measured in (a,c) $\mathbf{H} // \mathbf{ab}$ and (b,d) $\mathbf{H} // \mathbf{c}$ around $T_c$ for Cr$_2$X$_2$Te$_6$ (X = Si and Ge).}
\label{Arrot}
\end{figure}

\subsection{Magnetocrystalline anisotropy}

Figure 3 shows the magnetization isotherms with field up to 50 kOe applied for both $\mathbf{H//ab}$ and $\mathbf{H//c}$ around $T_c$ for Cr$_2$X$_2$Te$_6$ (X = Si and Ge). When $\mathbf{H//ab}$, the saturation field $H_{s}$ is associated with the uniaxial magnetocrystalline anisotropy parameter $K_u$ and the saturation magnetization $M_s$, i.e., $2K_u/M_s = \mu_0H_{s}$, where $\mu_0$ is the vacuum permeability.\cite{Cullity} The temperature dependence of $K_u$ as well as $M_s$ and $H_s$ for Cr$_2$X$_2$Te$_6$ (X = Si and Ge) are depicted in Fig. 4. The calculated $K_u$ for Cr$_2$Si$_2$Te$_6$ is about 61 kJ/m$^3$ at 5 K. It gradually decreases to 38 kJ/m$^3$ at $T_c$ = 32 K, comparable with the $K_u$ values in CrBr$_3$.\cite{Dillon} The anisotropy parameter $K_u$ is much lower for Cr$_2$Ge$_2$Te$_6$: about 12 kJ/m$^3$ at 44 K and 5.6 kJ/m$^3$ at $T_c$ = 63 K. For clarity, only the $K_u$ values from $T_c$ to 44 K are presented for Cr$_2$Ge$_2$Te$_6$. The $K_u$ of 20 kJ/m$^3$ at 2 K for Cr$_2$Ge$_2$Te$_6$ estimated from Fig. 2(b) is much smaller than that of 65 kJ/m$^3$ for Cr$_2$Si$_2$Te$_6$, in line with the magnetization data. The observed decrease of $K_u$ with increasing temperature arises solely from a large number of local spin clusters fluctuating randomly around the macroscopic magnetization vector and activated by a nonzero thermal energy, whereas the anisotropy constants are temperature-independent.\cite{Zener,Carr} This provides insight into the understanding of FM in few-layers of Cr$_2$X$_2$Te$_6$ (X = Si and Ge). While in pure 2D system no long-range magnetic order is expected,\cite{Mermin} mechanical corrugations and magnetic anisotropy are possible pathways to establish magnetism in few-layers samples.

\begin{figure}
\centerline{\includegraphics[scale=1]{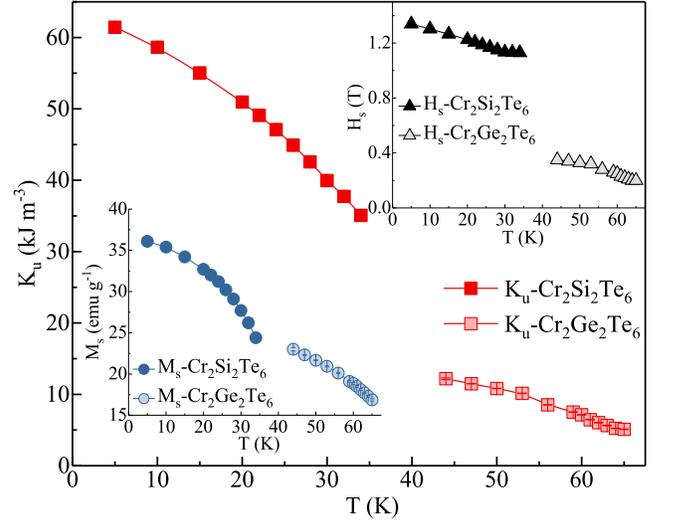}}
\caption{(Color online) Temperature dependence of the calculated anisotropy constant $K_u$, the estimated saturation field $H_s$ and the saturation magnetization $M_s$ (insets) below $T_c$ for Cr$_2$X$_2$Te$_6$ (X = Si and Ge).}
\label{renomalized}
\end{figure}

\begin{figure}o
\centerline{\includegraphics[scale=1]{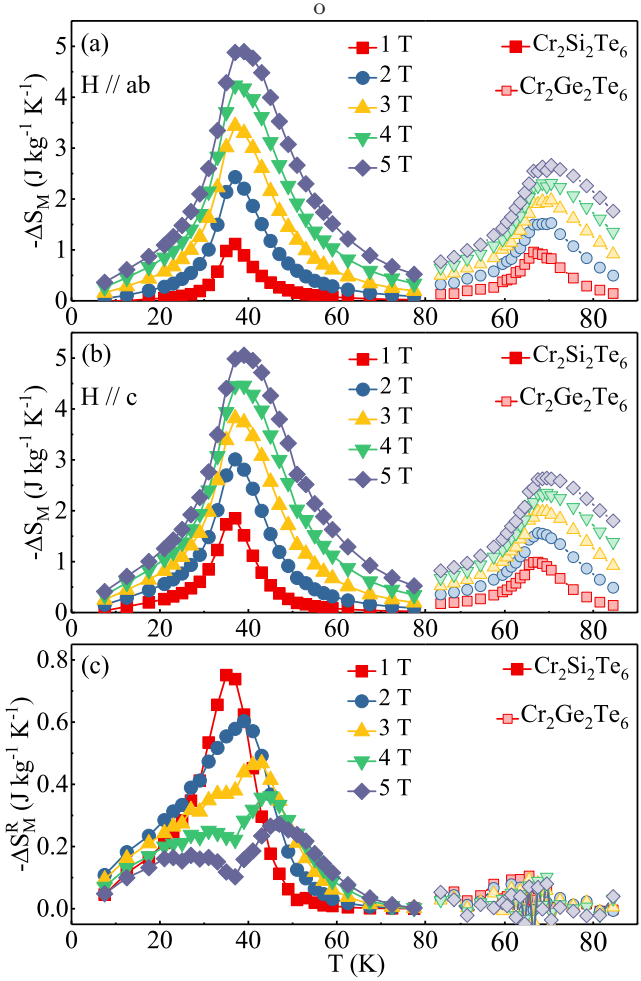}}
\caption{(Color online) Temperature dependence of isothermal magnetic entropy change $-\Delta S_M$ obtained from magnetization at various magnetic fields change (a) in the $ab$ plane and (b) along the $c$ axis, respectively, for Cr$_2$X$_2$Te$_6$ (X = Si and Ge). (c) Temperature dependence of $-\Delta S_M^R$ obtained by rotating from the $ab$ plane to the $c$ axis in various fields.}
\label{KF}
\end{figure}

\begin{figure}
\centerline{\includegraphics[scale=1]{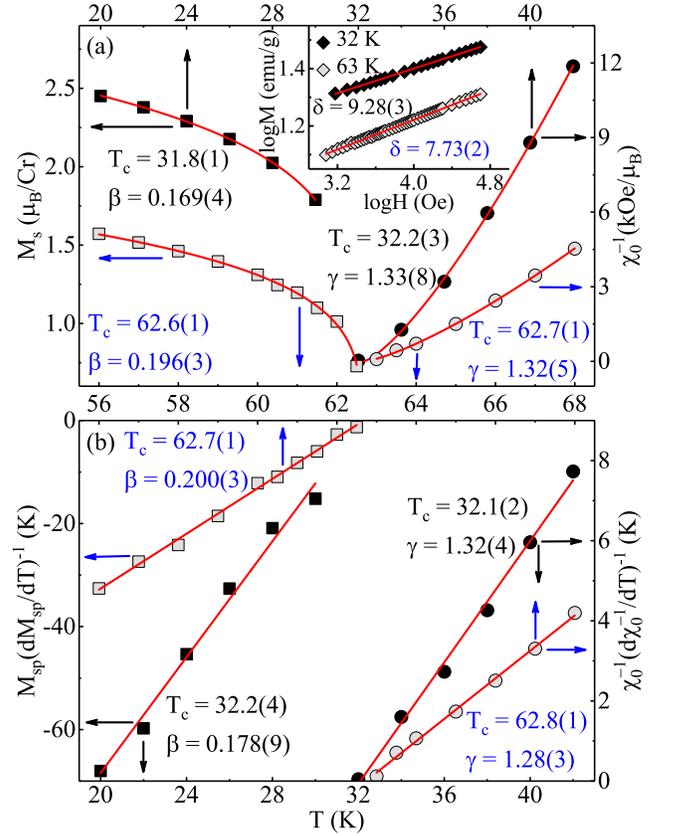}}
\caption{(Color online) (a) Temperature dependence of the spontaneous magnetization $M_{sp}$ (left axis) and the inverse initial susceptibility $\chi_0^{-1}$ (right axis) in out-of-plane field with solid fitting curves for Cr$_2$X$_2$Te$_6$ (X = Si and Ge). Inset shows log$M$ vs log$H$ collected at $T_c$ with linear fitting curves. (b) Kouvel-Fisher plots of $M_{sp}(dM_{sp}/dT)^{-1}$ (left axis) and $\chi_0^{-1}(d\chi_0^{-1}/dT)^{-1}$ (right axis) with solid fitting curves.}
\label{renomalized}
\end{figure}

\subsection{Magnetic entropy change}

We estimate the magnetic entropy change
\begin{equation}
\Delta S_M(T,H) = \int_0^H \left(\frac{\partial S}{\partial H}\right)_TdH = \int_0^H \left(\frac{\partial M}{\partial T}\right)_HdH,
\end{equation}
where $\left(\frac{\partial S}{\partial H}\right)_T$ = $\left(\frac{\partial M}{\partial T}\right)_H$ is based on Maxwell's relation. In the case of magnetization measured at small discrete field and temperature intervals [Fig. 3], $\Delta S_M$ can be approximated:
\begin{equation}
\Delta S_M(T_i,H) = \frac{\int_0^HM(T_i,H)dH-\int_0^HM(T_{i+1},H)dH}{T_i-T_{i+1}}.
\end{equation}
Figures 5(a) and 5(b) present the calculated $-\Delta S_M(T,H)$ as a function of temperature with in-plane and out-of-plane fields. All the $-\Delta S_M(T,H)$ curves show a pronounced peak at $T_c$, and the peak broads asymmetrically on both sides with increasing field. The maximum value of $-\Delta S_M$ is 4.9 J kg$^{-1}$ K$^{-1}$ for Cr$_2$Si$_2$Te$_6$ and 2.6 J kg$^{-1}$ K$^{-1}$ for Cr$_2$Ge$_2$Te$_6$ with in-plane field change of 5 T. These slightly increase to 5.05 and 2.64 J kg$^{-1}$ K$^{-1}$, respectively, with out-of-plane field change of 5 T. The obtained $-\Delta S_M$ values are comparable and larger than that of Fe$_3$GeTe$_2$ and CrI$_3$.\cite{Verchenko,YuLIU}
The rotational magnetic entropy change $\Delta S_M^R$ is calculated as $\Delta S_M^R(T,H) = \Delta S_M(T,H_c)-\Delta S_M(T,H_{ab})$. As shown in Fig. 5(c), the value of $\Delta S_M^R$ for Cr$_2$Si$_2$Te$_6$ is larger than that for Cr$_2$Ge$_2$Te$_6$, in line with the calculated $K_u$ [Fig. 4]. The anisotropy is gradually suppressed in higher field, and interestingly, it splits into two peaks on both sides of $T_c$ with field above 3 T for Cr$_2$Si$_2$Te$_6$.

\subsection{Critical behavior}

According to the scaling hypothesis, the second-order phase transition around $T_c$ can be characterized by a set of interrelated critical exponents and magnetic equation of state.\cite{Stanley} The exponents $\beta$ and $\gamma$ can be obtained from spontaneous magnetization $M_{sp}$ and inverse initial susceptibility $\chi_0^{-1}$, below and above $T_c$, respectively, while $\delta$ is a critical isotherm exponent at $T_c$. The mathematical definitions of the exponents from magnetization measurement are given below:
\begin{equation}
M_{sp} (T) = M_0(-\varepsilon)^\beta, \varepsilon < 0, T < T_c,
\end{equation}
\begin{equation}
\chi_0^{-1} (T) = (h_0/m_0)\varepsilon^\gamma, \varepsilon > 0, T > T_c,
\end{equation}
\begin{equation}
M = DH^{1/\delta}, T = T_c,
\end{equation}
where $\varepsilon = (T-T_c)/T_c$ is the reduced temperature, and $M_0$, $h_0/m_0$ and $D$ are the critical amplitudes.\cite{Fisher}

The critical exponents $\beta$, $\gamma$, and $\delta$, as well as the precise $T_c$ can be obtained by the modified Arrott plot of $M^{1/\beta}$ vs $(H/M)^{1/\gamma}$ in the vicinity of $T_c$ with a self-consistent method.\cite{Kellner, Pramanik} This gives $\chi_0^{-1}(T)$ and $M_{sp}(T)$ as the intercepts on the $H/M$ axis and the positive $M^2$ axis, respectively. Figure 6(a) presents the final $M_{sp}(T)$ and $\chi_0^{-1}(T)$ as a function of temperature. According to Eqs. (3) and (4), the critical exponents $\beta = 0.169(4)$ with $T_c = 31.8(1)$ K [$\beta = 0.196(3)$ with $T_c = 62.6(1)$ K], and $\gamma = 1.33(8)$ with $T_c = 32.2(3)$ K [$\gamma = 1.32(5)$ with $T_c = 62.7(1)$ K], are obtained for Cr$_2$Si$_2$Te$_6$ [Cr$_2$Ge$_2$Te$_6$].

Based on the Kouvel-Fisher (KF) relation:\cite{Kouvel}
\begin{equation}
M_{sp}(T)[dM_{sp}(T)/dT]^{-1} = (T-T_c)/\beta,
\end{equation}
\begin{equation}
\chi_0^{-1}(T)[d\chi_0^{-1}(T)/dT]^{-1} = (T-T_c)/\gamma.
\end{equation}
Linear fitting to the plots of $M_{sp}(T)[dM_{sp}(T)/dT]^{-1}$ vs $T$ and $\chi_0^{-1}(T)[d\chi_0^{-1}(T)/dT]^{-1}$ vs $T$, as shown in Fig. 6(b), yield $\beta = 0.178(9)$ with $T_c = 32.2(4)$ K [$\beta = 0.200(3)$ with $T_c = 62.7(1)$ K], and $\gamma = 1.32(4)$ with $T_c = 32.1(2)$ K [$\gamma = 1.28(3)$ with $T_c = 62.8(1)$ K]. The third exponent $\delta$ can be calculated from the Widom scaling relation $\delta = 1+\gamma/\beta$. From $\beta$ and $\gamma$ obtained with the modified Arrott plot and the Kouvel-Fisher plot, $\delta$ = 8.9(3) and 8.4(2) [7.7(2) and 7.4(1)] for Cr$_2$Si$_2$Te$_6$ [Cr$_2$Ge$_2$Te$_6$], which are close to the direct fits of $\delta$ taking into account that $M = DH^{1/\delta}$ at $T_c$ [$\delta$ = 9.28(3) at 32 K and 7.73(2) at 63 K, inset in Fig. 6(a)]. The critical exponents of Cr$_2$X$_2$Te$_6$ (X = Si and Ge) are summarized in Table II. They are close to but not identical to values expected for the 2D-Ising model ($\beta$ = 0.125, $\gamma$ = 1.75 and $\delta$ = 15). This deviation is most likely associated with non-negligible interlayer coupling and spin-lattice coupling in this system.\cite{Casto, Carteaux3}

\begin{figure}
\centerline{\includegraphics[scale=1]{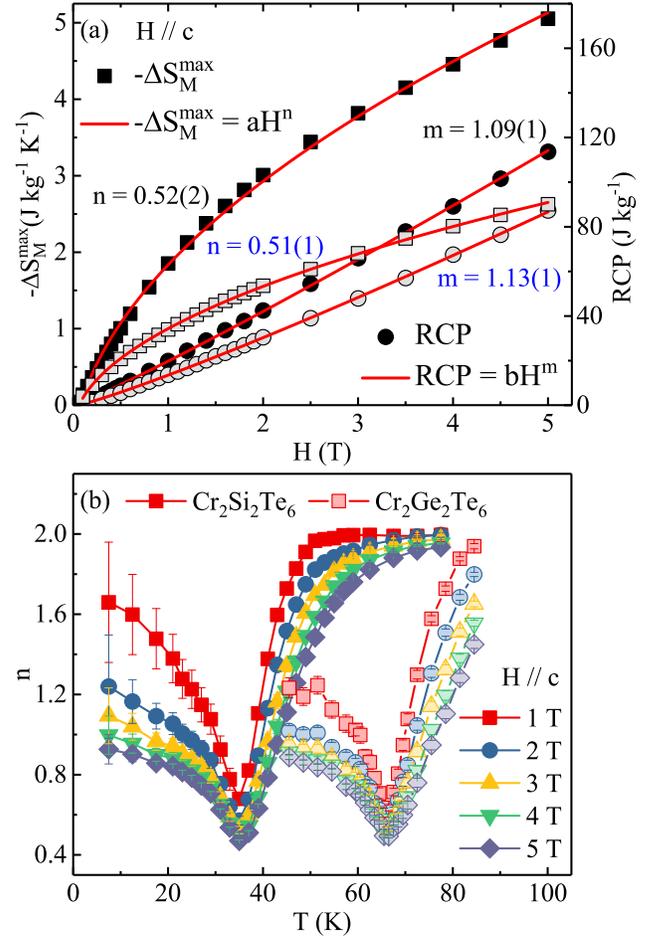}}
\caption{(Color online) (a) Field dependence of the maximum magnetic entropy change $-\Delta S^{max}_M$ and the relative cooling
power $RCP$ with power law fitting in red solid lines for Cr$_2$X$_2$Te$_6$ (X = Si and Ge). (b) Temperature dependence of $n$ in various fields.}
\label{renomalized}
\end{figure}

\subsection{Scaling analysis of the $\Delta S_M$ data}

\begin{table}
\caption{\label{tab}Critical exponents of Cr$_2$X$_2$Te$_6$ (X = Si and Ge). The MAP, KFP and CI represent the modified Arrott plot, the Kouvel-Fisher plot and the critical isotherm, respectively.}
\begin{ruledtabular}
\begin{tabular}{lllllll}
   & & $\beta$ & $\gamma$ & $\delta$ & $n$ & $m$\\
   \hline
   X = Si & & & & & & \\
   & $-\Delta S_M^{max}$  &   &   &   & 0.52(2) & \\
   & RCP  &   &   &   &   & 1.09(1)\\
   & MAP & 0.169(4) & 1.33(8) & 8.9(3) & 0.45(3) & 1.112(4) \\
   & KFP & 0.178(9) & 1.32(4) & 8.4(2) & 0.45(3) & 1.119(3) \\
   & CI  &   &   & 9.28(3) & & 1.108(1) \\
   \hline
   X = Ge & & & & & & \\
   & $-\Delta S_M^{max}$  &   &   &   & 0.51(1) & \\
   & RCP  &   &   &   &   & 1.13(1)\\
   & MAP & 0.196(3) & 1.32(5) & 7.7(2) & 0.47(2) & 1.130(3) \\
   & KFP & 0.200(3) & 1.28(3) & 7.4(1) & 0.46(1) & 1.135(2) \\
   & CI  &   &   & 7.73(2) & & 1.129(1) \\
\end{tabular}
\end{ruledtabular}
\end{table}

\begin{figure}
\centerline{\includegraphics[scale=1]{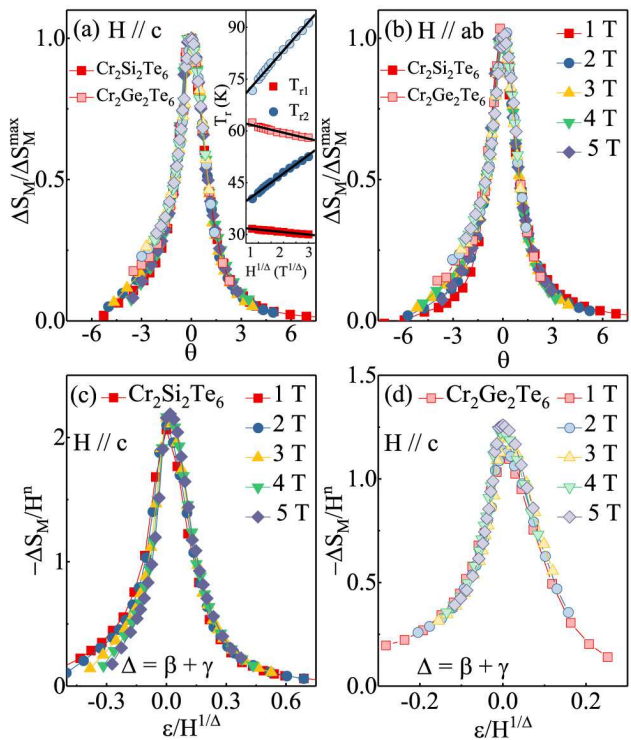}}
\caption{(Color online) The normalized $\Delta S_M$ as a function of the reduced temperature $\theta$ with (a) out-of-plane and (b) in-plane field for Cr$_2$X$_2$Te$_6$ (X = Si and Ge). Scaling plot for (c) Cr$_2$Si$_2$Te$_6$ and (d) Cr$_2$Ge$_2$Te$_6$ based on the critical exponents $\beta$ and $\gamma$ obtained in out-of-plane field.}
\label{renomalized}
\end{figure}

For a material displaying a second-order transition,\cite{Oes} the field dependence of the maximum magnetic entropy change shows a power law $-\Delta S_M^{max} = aH^n$,\cite{VFranco} where $a$ is a constant and the exponent $n$ at $T_c$ is related to the critical exponents as $n(T_c) = 1+(\beta-1)/(\beta+\gamma)$. Another important parameter is the relative cooling power (RCP): $RCP = -\Delta S_M^{max} \times \delta T_{FWHM}$  where $-\Delta S_M^{max}$ is the maximum entropy change near $T_c$ and $\delta T_{FWHM}$ is the full-width at half maximum.\cite{Gschneidner} The RCP also depends on the magnetic field with $RCP = bH^m$, where $b$ is a constant and $m$ is associated with the critical exponent $\delta$, $m = 1+1/\delta$.

Figure 7(a) presents the summary of the out-of-plane field dependence of $-\Delta S_M^{max}$ and RCP. The calculated values of RCP are about 114 and 87 J kg$^{-1}$ for Cr$_2$Si$_2$Te$_6$ and Cr$_2$Ge$_2$Te$_6$, respectively, with out-of-plane field change of 5 T. Fitting of the $-\Delta S_M^{max}$ gives that $n = 0.52(2)$ and 0.51(1) for Cr$_2$Si$_2$Te$_6$ and Cr$_2$Ge$_2$Te$_6$, respectively [Fig. 7(a)], which is quite close to that of $n = 0.53$ for the 2D-Ising model ($\beta = 0.125$, $\gamma = 1.75$). Fitting of the RCP generates $m = 1.09(1)$ and 1.13(1) [Fig. 7(a)], which is also close to the expected value of 1.07 for 2D-Ising model ($\delta$ = 15). Figure 7(b) displays the temperature dependence of $n(T)$ in various fields. All $n(T)$ curves follow an universal behavior.\cite{FrancoV} At low temperatures, well below $T_c$, $n$ has a value around 1. On the other side, well above $T_c$, $n$ is close to 2 as a consequence of the Curie-Weiss law. At $T = T_c$, $n(T)$ has a minimum.

Scaling analysis of $-\Delta S_M$ can be built by normalizing all the $-\Delta S_M$ curves against the respective maximum $-\Delta S_M^{max}$, namely, $\Delta S_M/\Delta S_M^{max}$ by rescaling the reduced temperature $\theta_\pm$ as defined in the following equations,\cite{Franco}
\begin{equation}
\theta_- = (T_{peak}-T)/(T_{r1}-T_{peak}), T<T_{peak},
\end{equation}
\begin{equation}
\theta_+ = (T-T_{peak})/(T_{r2}-T_{peak}), T>T_{peak},
\end{equation}
where $T_{r1}$ and $T_{r2}$ are the temperatures of two reference points that corresponds to $\Delta S_M(T_{r1},T_{r2}) = \frac{1}{2}\Delta S_M^{max}$. Following this method, all the $-\Delta S_M(T,H)$ curves in various fields collapse into a single curve in the vicinity of $T_c$ for Cr$_2$X$_2$Te$_6$ (X = Si and Ge), as shown in Figs. 8(a) and 8(b). The values of $T_{r1}$ and $T_{r2}$ depend on $H^{1/\Delta}$ with $\Delta = \beta + \gamma$ [inset in Fig. 8(a)].

In the phase transition region, the scaling analysis of $-\Delta S_M$ can also be expressed as
\begin{equation}
\frac{-\Delta S_M}{a_M} = H^nf(\frac{\varepsilon}{H^{1/\Delta}}),
\end{equation}
where $a_M = T_c^{-1}A^{\delta+1}B$ with A and B representing the critical amplitudes as in $M_{sp}(T) = A(-\varepsilon)^\beta$ and $H = BM^\delta$, respectively, and $f(x)$ is the scaling function.\cite{Su} If the critical exponents are appropriately chosen, the $-\Delta S_M$ vs $T$ curves should be rescaled into a single curve, consistent with normalizing the $-\Delta S_M$ curves with two reference temperatures ($T_{r1}$ and $T_{r2}$). As shown in Figs. 8(c) and 8(d), the rescaled $-\Delta S_M$ for Cr$_2$X$_2$Te$_6$ (X = Si and Ge) with out-of-plane field collapse onto a single curve, confirming the reliable critical exponents and 2D Ising behavior for Cr$_2$X$_2$Te$_6$ (X = Si and Ge).

\section{CONCLUSIONS}

In summary, we have studied the critical behavior and magnetocaloric effect around the FM-PM transition in Cr$_2$X$_2$Te$_6$ (X = Si and Ge) single crystals. The critical exponents $\beta$, $\gamma$, and $\delta$ estimated from various techniques match reasonably well and the scaling analysis of magnetic entropy change confirms that they are 2D Ising ferromagnents with non-negligible interlayer coupling. The uniaxial magnetocrystalline anisotropy confirmed here could be the possible origin of existence of long-range FM in few-layers of Cr$_2$X$_2$Te$_6$ (X = Si and Ge).

\section*{Acknowledgements}
We thank Mark Dean for useful discussions. Work at Brookhaven is supported by the Research supported by the U.S. Department of Energy, Office of Basic Energy Sciences as part of the Computation Material Science Program (Y. L. and C. P.) and by the US DOE under Contract No. DE-SC0012704 (C. P.).

\end{document}